Perception and recognition in a neural network theory in which neurons exhibit hysteresis

Geoffrey W. Hoffmann
Department of Physics and Astronomy
University of British Columbia
Vancouver, B.C.
Canada V6T 1Z7

**A neural network theory of visual perception and recognition is presented. Information flows both from the retina to the brain and from the brain to the retina[1]. A report that when a scene is perceived 50 retinal cells are much more active than any of the other retinal cells is ascribed significance in the theory[2]. The theory involves neurons that exhibit hysteresis, without the need for any changes in synaptic connection strengths during learning[3]. The fact that the brain is able to recognize faces and other objects very rapidly[4] is discussed in the context of the theory. The theory can be epitomized as "We see with our eyes and remember with our brains".**

David Hubel describes the predominant neuroscience paradigm in which signals from sensory organs proceed through successive layers of neurons, going deeper and deeper into the brain[1]. For example, in the case of vision signals go from the retina, through the optic nerve, to the lateral geniculate body, to the cortex, and so on. He also writes, however, that "We need to qualify our model with respect to the direction of information flow. The prevailing direction in our diagram on page 24 is obviously from left to right, from input to output, but in almost every case in which information is transferred from one stage to the next, reciprocal connections feed information back from the second stage to the first. (We can sometimes guess what such feedback might be useful for, but in almost no case do we have incisive understanding.) Finally, even within a given stage we often find a rich network of connections between neighboring cells of the same order. Thus to say that a structure contains a specific number of stages is almost always an oversimplification."

John Hopfield has written that "When you look at a particular person there will be 50 cells in the retina which are much more active than any of the other retinal cells. If the lighting is changed a little, or if a highlight moves somewhere else in the scene, or if the person moves, the retinal cells which are most strongly driven completely change. The 50 most intensely driven cells actually say nothing very useful about the nature of objects 'out there'"[2].

These observations can be interpreted in the context of a neural network model in which neurons exhibit hysteresis[3]. In the interpretation the retinal cells are influenced both by the external light stimulus being experienced and on signals coming from the brain to the retina. The crux of the model is that memory is the domain of the brain, while the site of immediate visual perception is the retina. Most simply put, we remember with our brains and see with our eyes. The perceived image is then recognized by the brain.

The idea is that the set of 50 highly active retinal cells depends on our set of memories. An important question is then how much information can be contained in the selection of the 50 cells from the 100,000,000 retina cells. The number of such combinations is $\frac{10^8!}{(10^8-50)!50!} \approx 10^{335}$. This number of combinations is plausibly sufficient to account for the number of visual memory states of an individual. If the selections of 50 retinal cells correlate with visual memories, the fact that a small change in input can cause a complete change in the 50 highly driven cells means that many quite different selected sets correlate with similar sets of visual memories. A one-to-one correlation of the selected set of 50 retinal cells with the recognition of particular persons or objects is not necessary and is not envisaged, since our complete set of memories is developing on an ongoing basis. On the other hand it is envisaged that when a person or thing is recognized there is a strong correlation in the vector of activation of the $10^{11}$ neurons of the entire human brain and the corresponding vector the last time that person or object was recognized. Visual perception is by the eyes and recognition is by the brain.

The visual perception of a scene is postulated to correlate with low levels of activation of the remaining $10^8 - 50$ retinal cells, below the level that would result in switching from the low arm of a hysteresis curve to the high arm. Perceptual visual information is then contained in the precise levels of activation, which would vary across the retina according to the level of light, without causing the retinal cells to switch to the high level of activation. These levels of activation constitute $10^8 - 50$ continuous variables, each with, in principle, up to an infinite number of possible values. Hence, again in principle, the perceived image has of the order of $\infty^{100,000,000}$ possible states. This is many more than needed to account for visual perception. Each of these patterns of excitation, or visual perception, occurs in the context of the selection of 50 highly activated retinal cells, corresponding to one of the up to $10^{335}$ theoretical visual memory states. Perception of a familiar face involves the selection of 50 highly driven retinal cells, namely one of the sets that correlates with the eyes being exposed to that recognized face. The selection of the 50 cells is a function of both the pattern of light on the retina and the set of stimuli arriving at the retina from the brain.

A problem for neural network theory has been the rapidity with which the brain is able to recognize people and other objects[4]. The model has the potential to account for this phenomenon, since the place where visual perception occurs is envisaged as being at the retina itself. A complete set of visual memories of the person is present there, encoded in the context of all the fixed synaptic connections of the brain as the selected set of 50 highly driven retinal cells. On presentation of an image, there is a convergence at the retina of information from the external stimulus and from the brain. When we see a familiar face, we can envisage that the convergence results in both perception and recognition. Perception is viewed as the formation of a pattern of low level activity at the retina, while recognition involves a shift in the states of most of the $10^{11}$ neurons of the brain closer to a familiar region of the $10^{11}$-dimensional phase space, corresponding to a remembered object. The perception step involves feedback between the brain (memory) and the retina (visual input), and is consequently expected to be rapid for a familiar image. The recognition step is also rapid, and corresponds to the state vector for all the neurons of the brain being perturbed in a way that is correlated with perturbations of the state vector that occurred at times of previous presentations with the same visual stimulus. Rapid perception is thus understood in this model as being the result of the retina being driven by the pattern of light in the context of the set of memories of the brain, as reflected in the set of highly driven cells. Perception of an image includes the activation of one of the sets of highly driven cells in the retina that correlate with the recognition of the individual. Seeing the retina as the site where visual perception occurs solves the problem that there is no known mechanism for constructing a sharp two-dimensional image deep in the brain.

Exposures to unfamiliar faces are expected to not evoke the activation of a narrowly defined set of 50 neurons. Multiple exposures to a new face will however gradually lead to learning that face, with systematic narrowing in the selection of the highly driven retinal cells whenever that face is presented, and the development of a new familiar region of the phase space, as defined by the rates of firing of all the neurons of the brain. This establishes the basis for the subsequent recognition of the face.